# CONDUCTANCE OF A SEMICONDUCTOR(2DEG) - SUPERCONDUCTOR JUNCTION IN HIGH MAGNETIC FIELD

N.M. Chtchelkatchev[*]

[*]L.D.Landau Institute for Theoretical Physics RAS, 117940 Moscow, Russia

e-mail: nms@landau.ac.ru

Conductance $G$ of a 2DEG-Superconductor (S) device in a high magnetic field is studied: $G(\nu)$ is calculated. When the cyclotron diameter in 2DEG is larger than the width of the 2DEG-S surface then $G(\nu)$ becomes nonmonotonous function due to the Aharonov–Bohm type interference of quasiparticles at the surface. At certain parameters of the junction the conductance oscillates with $\nu$.

In recent years, the study of hybrid systems consisting of superconductors in contact with normal metals in strong magnetic field has attracted considerable interest [1] - [5]. Investigation of physical phenomena in S-2DEG devices in high magnetic field may help to establish a link between mesoscopic superconductivity and quantum - Hall physics. It was found experimentally [1] that zero-bias conductance $G$ of a ballistic S - 2DEG - S junction in Integer Quantum Hall (IQH) regime exhibits quantization under variation of magnetic field. The quantum of $G$ was not equal to a universal value in this experiment, as for instance in IQH or in a quantum point contact [7], but it was an oscillating function of the field $H$. Numerical simulations [4], [5] showed that the conductance of a 2DEG-S contact in IQH regime is a nonmonotonous function of the filling factor $\nu$; there is nonuniversal quantization of $G$ when 2DEG-S boundary is perfect [6]; at specific range of magnetic field $G(\nu)$ oscillates. A phenomenological theory of the conductance oscillations was suggested in [5]. But, it is still unclear when the conductance becomes sensitive to $H$, why it exhibits oscillations, how one can analytically describe $G(H)$. The analytical form of $G(\nu)$ is found in this paper. It is shown that the conductance becomes sensitive to $H$ when $2R_c \gtrsim L$, where $R_c$ is a cyclotron radius in 2DEG, $L$ characterizes the length of the 2DEG-S boundary; nonlinearities of $G(\nu)$ result from Aharonov–Bohm type interference of quasiparticles at the boundary.

We consider a junction consisting of a superconductor, 2DEG and a normal conductor segments (see Fig.1). Magnetic field $H$ is applied along $z$ direction, perpendicular to the plain of 2DEG. It is supposed that quasiparticle transport is ballistic (the mean free path of an electron $l_{tr} \gg L$, where $L$ is the length of the 2DEG-S boundary). The current $I$ is supposed to flow between normal (N) and superconducting (S) terminals (the voltage $V$ is applied between them). The conductance $G(H, L) = I/V, V \to 0$ is studied in the paper.

Following [8], we shall describe transport properties of the junction in terms of electron and hole quasiparticle scattering states, which satisfy Bogoliubov-de Gennes (BdG) equations. Then the conductance

$$G = \left.\frac{\partial I}{\partial V}\right|_{V \to 0} = \frac{2e^2}{h} \sum_{l_o, n_i} R_{he, l_o n_i} = \frac{4e^2}{h} \mathcal{R}, \tag{1}$$

where $R_{he,l_o n_i}$ is probability of Andreev reflection of an electron with the energy $E = 0$ (with the respect to $E_f$) incident on the superconductor in the channel with quantum number $n_i$ to a hole going from the superconductor in the channel $l_o$. Before explicit

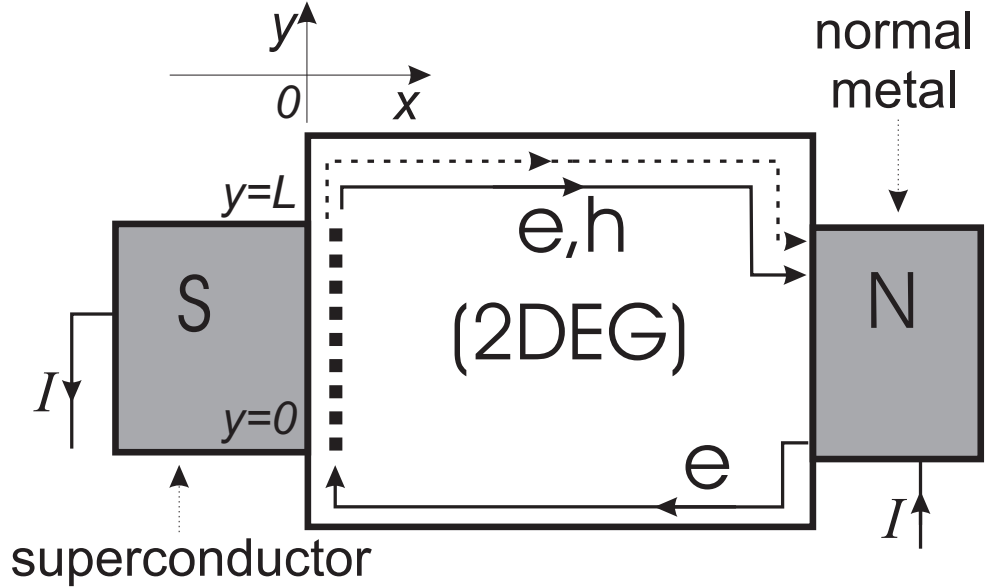

. 1. The device, which we investigate, consists of a superconductor, 2DEG and a normal conductor. An electron injected from the normal conductor in IQH regime goes through an edge state to the superconductor, reflects into a hole and an electron which return to the normal contact through the other edge states.

calculation of $G$, we shall discuss on a qualitative level how $G$ should depend on $H$. When $H$ is small ($R_c \ll L$) then $R_{he,l_o n_i} \simeq R_{he} \delta_{l_o,n_i}$, with $R_{he}$ weakly depending on $H$. So,

$$\mathcal{R} \simeq R_{he} N, \ N = L p_f / 2\pi. \qquad (2)$$

If $2R_c \gtrsim L$, quasiparticles reflected from the superconductor (S) due to normal and Andreev reflection of the electron return to S again due to bending of the trajectories by magnetic field. Then $G(H)$ dependence is not weak. We shall investigate the conductance using semiclassical approximation when $\nu \gg 1$. An electron (hole) quasiparticle in 2DEG can be viewed in semiclassics as a beam of rays (in a similar way propagation of light is described by beams of rays in classical optics [9]). Trajectories of the quasiparticle rays can be found from the equations of classical mechanics. If $R_c \gg L$ ($\nu \ll L/\lambda_f$), the edge states at 2DEG do not overlap. Then the quantum numbers $n_o, l_i$ (1) of the incident electron and reflected hole correspond to the edge states. Reflection of an electron from the superconductor is schematically shown in Fig.1. The electron ray (Fig.2) reflects into electron and hole rays from S at $y_0$. These rays reflect into other hole and electron rays at $y_1$. So, two hole and two electron rays propagate between $y_2$ and $y_3$. Then eight hole rays come from S to N propagating along the same hole path beginning at $y_3$. One can

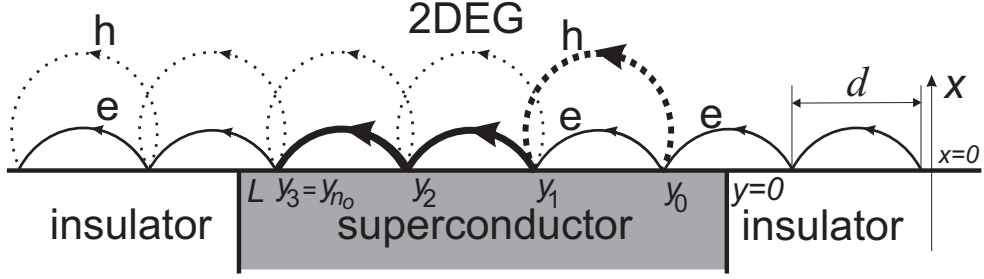

. 2. A quasiparticle wave in quasiclassical approximation can be treated as a beam of rays. The figure displays reflection of a ray (solid curve), corresponding to an electron injected from N, from the superconductor. Dashed curves correspond to hole rays.

approximate the probability $P$ of this hole path by the expression:

$$P(y_0, n_i, n_o) \simeq \left| r_{he} r_{ee} r_{ee} r_{ee} e^{3iS_e - i3\pi/2} + r_{hh} r_{he} r_{ee} r_{ee} e^{iS_h + 2iS_e - i\pi/2} + \ldots \right|^2, \qquad (3)$$

where $r_{ba}$ is the amplitude of reflection of a quasiparticle $a$ into a quasiparticle $b$ from the superconductor. $S_{e(h)}$ is the quasiclassical action of an electron (hole) taken along the part of the trajectory connecting adjacent points of reflection. Then $R_{eh,l_o n_i} \simeq \langle P(y_0)\rangle \delta_{n_i,l_0}$, where the average is taken over $0 < y_0 < d(n_i)$, with $d$ being the length of the quasiparticle "jump" along the edge of 2DEG (see Fig. 2). Expression (3) includes interference terms which depend on $S_e - S_h$. As $S_e - S_h = 2\pi(\nu - 1/2)$, one can expect the conductance to be a nonlinear function of $\nu$ due to the interference terms. The nature of this nonlinearity indeed resembles Aharonov–Bohm effect, as it was supposed in [5], where the conductance oscillates with $H$ because vector potential changes phases of electrons going from source to drain along different paths. It will be seen below that at certain conditions S-2DEG conductance oscillates with $\nu$.

Semiclassical estimates used above supposed that there is spin degeneracy; $T, eV = 0$; diffraction is small: difference of hole and electron momentum at $E > 0$ was neglected. This approximation is valid when $\max\{T, |eV|, g\mu_B H\}/\mu \ll \lambda_F/L$, where $\lambda_F$ is Fermi wavelength in 2DEG. Calculation of the conductance below also supposes these conditions to be satisfied.

The conductance of 2DEG-S structure will be calculated below as a semiclassical asymptotic ($\nu \gg 1$) of (1). If the S-2DEG surface is flat then the approach of [8], [10] gives an idea how one can express $R_{eh,ln}$ via semiclassical asymptotic of greens functions of BdG equations. Doing this calculation we confirm, that above naive estimates of $G$ really lead to semiclassical asymptotic of the conductance:

$$\mathcal{R} = \sum_{n_i} \int_0^{d(n_i)} \left\{ \rho(n_i, y_0) \left| \sum_a t_a \exp\left(iS_a - i\frac{\pi}{2}\mu_a\right) \right|^2 \right\} dy_0, \qquad (4)$$

where $n_i$ is the index of an edge state of an electron incident on the superconductor, $d(n_i)$ is the length of a quasiparticle jump; $t_a$ is the probability amplitude for the classical quasiparticle trajectory from $y = y_0$ to $y = y_{n_o}$ – the coordinate of the last reflection from the superconductor. The amplitude $t_a$ is a product of Andreev and normal reflection amplitudes; $S_a$ is the action taken from $y_0$ – the coordinate of the first quasiparticle reflection to $y_{n_o}$ – the coordinate of the last reflection (see Fig. 2); $\mu_a$ is Maslov index of the trajectory. For example, $t_a = r_{he} r_{ee} r_{eh} r_{he}$ and $S_a = S_h + 2S_e$ for the trajectory distinguished by a thick line in Fig. 2. Summation over $a$ means the sum over all paths connecting $y_0$ with $y_{n_o}$ at 2DEG-S boundary. The wight function $\rho(n_i, y_0)$, where $\int_0^d \rho(n_i, y_0) dy_0 = 1$, generally depends on the shape of 2DEG-S contact. If 2DEG spreads over the region $x > 0$, $y > 2R_c + L$, $y < -2R_c$, as it is in Fig. 2, then $\rho = 1/d$. Formula (4) is a central result of the paper.

The sum over trajectories in (4) could be converted into an analytical expression:

$$G = \frac{4e^2}{h} \sum_{n_i} \sum_s P_s \frac{R_{eh} \sin^2(s \arccos(\sqrt{R_{ee}} \cos(\Omega)))}{1 - R_{ee} \cos^2(\Omega)}, \qquad (5)$$

where $\Omega = \pi \nu + \theta - 2\lambda p_\perp$; $\theta = arg(r_{ee})$ is the phase of the amplitude of electron – electron reflection from the superconductor, $R_{ee} = |r_{ee}|^2$ ; $p_\perp = p_\perp(n_i)$ is the perpendicular component of momentum of a quasiparticle when it reflects from the superconductor; $\lambda$ is equal to the penetration length of the superconductor; $\nu = E_f/(\hbar w_c) - 1/2$. Function $P_s$ is the probability to have $s$ reflections from the surface of the superconductor. When $\rho = 1/d$ this function could be expressed through the maximum number of jumps $g = [L/d]$ over the S-2DEG surface, where $[\ldots]$ denotes the integer part:

$$P_s = \begin{cases} \frac{L - gd}{d} & \text{if } s = g+1, \\ 1 - \frac{L-gd}{d} & \text{if } s = g, \\ 0 & \text{otherwise.} \end{cases} \qquad (6)$$

Expressions (4-5) are the central result of the paper; they show how the conductance depends on magnetic field and parameters of the contact. At small magnetic field (5) reduces to (2). If $2R_c \gtrsim L$, it follows from (5) that the conductance becomes sensitive to H. Few limiting cases of (5) will be considered below in this regime. The Aharonov–Bohm type conductance oscillations are the most interesting property of $G(\nu)$. It follows from (5) that oscillations are visible when $\lambda/L \ll R_c^2/L^2$ and $R_{eh} \lesssim 1/2$. A typical contact where S terminal is prepared with superconductor of first type, 2DEG formed in GaAs, has $\lambda \sim \lambda_F \sim 10^{-6} cm$, $L \sim 10^{-3} cm$. It follows from given above conditions that if $R_c \sim L$, oscillations can be seen in the contact. (These oscillations were numerically investigated in [4], [5]. It was checked that there is consistence between the theory presented in our paper and the numerical calculations.) When $R_{ee} \ll 1$ the conductance shows steps. It is interesting to investigate the regime $R_{eh} \ll 1$, $L/R_c \gg 1$. Then functional dependence $G(\nu)$ resembles light intensity distribution $I(\delta)$ seen in optics with Lummer - Gerike interferometer [9]. If one considers $s$ as the number of beams in the interferometer, $\delta = 2\Omega$ as the phase difference between successive beams. The probability $R_{he}$ will correspond to the transmission probability through the mirror of the interferometer. Examination of quasiparticle trajectories in (4), which give the main contribution to the conductance, shows that they are similar to trajectories of light beams in the interferometer.

Fig. 3 illustrates how the conductance depends on the magnetic field according to (5). One curve corresponds to $\lambda p_f = 1$ (dots), another – $\lambda p_f = 3$ (solid line). The solid line parallel to the X axes represents the conductance (2). If $2R_c \lesssim L$ the conductance

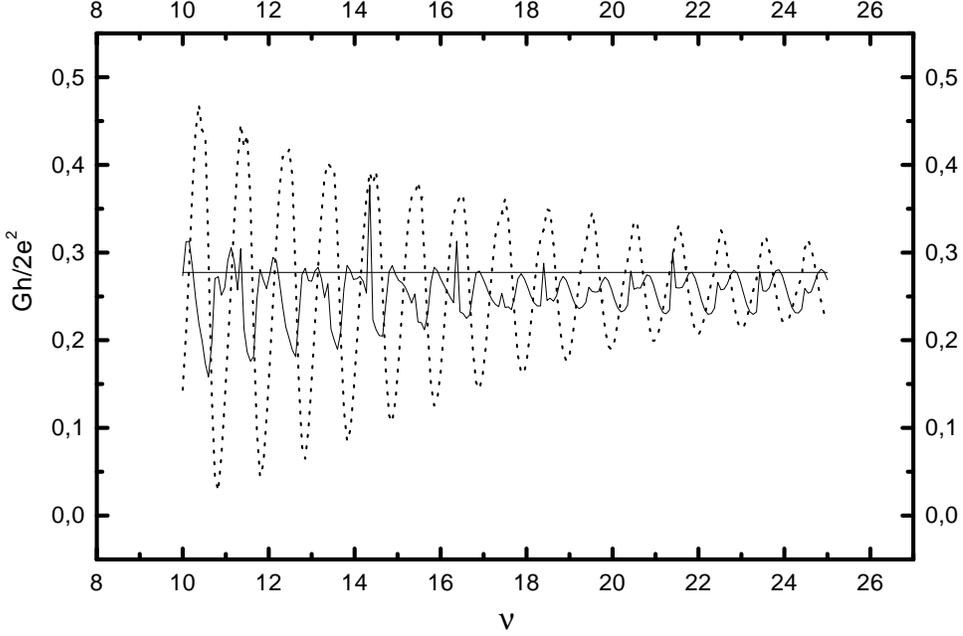

. 3. The curve drawn with the solid line represents the conductance for $\lambda L = 3$, with dots – for $\lambda L = 1$. The curves drawn in the figure approach $G(H = 0)$ (2) for large $\nu$.

oscillates because of quasiparticle interference, but when $\nu$ becomes larger ($2R_c \gtrsim L$) then interference phenomena become less probable and the conductance approaches (2). The graph also illustrates that increasing the ratio $\lambda/L$ leads to smearing of the oscillations. The following parameters of the contact were used: $Lp_f = 80$, $Z = 5$. Parameter $Z$ [11] characterizes normal scattering from the superconductor due to Shottki barriers, differences of the effective masses in 2DEG, S and so on. Amplitudes $r_{ee}, r_{eh}$ of normal and Andreev reflection from the superconductor were calculated for zero magnetic field by matching quasiparticle wave functions in 2DEG to the wave functions in S. This procedure is true while $(\lambda/R_c)^2 \ll 1$, where $R_c$ is a cyclotron radius in the superconductor. Magnetic fields used for making the plot satisfy this condition. The ratio $\Delta_0/E_f$ were equal to 0.02, with the gap $\Delta(x) = \Delta_0$ in the superconductor and zero in 2DEG.

It seems to be important to determine how a disorder at S-2DEG surface can influence on $G(\nu)$. The disorder could be represented by the roughness of the surface, impurities and so on. Disorder at the surface can induce fluctuations of $S_a$, $t_a$ in (4), break the interference of quasiparticles. We will characterize disorder by mean free path $l_{tr}$ of electron elastic scattering on impurities, by mean square root deviations $\delta n = \sqrt{\langle(\mathbf{n} - \mathbf{n_0})^2\rangle}$ of the normal unit vector to S-2DEG boundary from the direction $\mathbf{n_0}$ of the $X$ axes, where $\langle(\mathbf{n} - \mathbf{n_0})^2\rangle = \int_0^L dy(\mathbf{n}(y) - \mathbf{n_0})^2/L$. Then one can deduce that formula (5) is correct if $R_c \ll l_{tr}$ and $\delta n \ll \min\{l_H^2/L\lambda, 1/p_f L\}$.

When the surface of the superconductor is diffusive, i.e. $\delta n \gtrsim \max\{1/\lambda p_f, 1/\nu\}$ or $l_{tr} \lesssim R_c$, then there would be no interference between different paths in (4). One can estimate the reflection probability in this regime neglecting interference terms in (5).

Then

$$\mathcal{R} \simeq \sum_{n_i} \int_0^{d(n_i)} dy_0 \sum_a |t_a|^2 / d(n_i) \qquad (7)$$

The number of quasiparticle reflections from the surface of the superconductor would be about $s_0 = [L/2R_c]$ then

$$G = \frac{4e^2}{h} \mathcal{R} \simeq [\nu] \frac{4e^2}{\hbar} \langle R_{eh} \rangle (1 - 2\langle R_{eh} \rangle)^{1-s_0/2} U_{s_0}(\langle R_{ee} \rangle) / \sqrt{1 - 2\langle R_{eh} \rangle}, \qquad (8)$$

where $U_s(x) = \sin(s \arccos(x))/\sin(\arccos(x))$ is the Chebishev polinomial of the second kind [12]. When $s_0 \to \infty$ then the conductance (8) will approach $[\nu] 2e^2/h$.

One can suppose that the deviations from ideal conductance quantization (with the universal step $2e^2/h$) observed in the experiment [1] originate from the interference of quasiparticles studied above. The S-2DEG boundary of the device used in [1] was hardly flat, so contributions from harmonics with large $s$ (see eq. 5) should be suppressed. Then it is reasonable to approximate the conductance $G$ by $(2e^2/h)f(\nu)(1 + a\cos(\pi\nu + \varphi_0))$, where $a \ll 1$ characterizes disorder at the surface, $\varphi_0$ is a phase - shift, $f(\nu) \sim [\nu]$ describes the shape of the "quantum" of the conductance. It was checked that this formula is a good fit to the experimental data.

We thank G.B.Lesovik, M.V.Feigelman, Yu.V.Nazarov, A.Ioselevich, A.Shytov for helpful discussions. It is the Lesoviks interest that stimulated the appearance of the article. This work was supported by the Russian Foundation for Basic Research, project no. 00-02-16617. The author is grateful to The Netherlands Organization for Scientific Research (NWO) for support provided in the course of Dutch-Russian collaboration.

1. H.Takayanagi, T.Akazaki, Physica (Amsterdam) **249-251B**, 462 (1998).
2. T.D.Moore, D.A.Williams, Phys.Rev.B **59**, 7308 (1999)
3. H.Hoppe, U.Zülike, and G.Schön, Phys.Rev.Lett. **84**, 1804 (2000).
4. Y.Takagaki, Phys.Rev.B **57**, 4009 (1998).
5. Y.Asano, Phys.Rev.B **61**, 1732 (2000); Y.Asano, T.Yuito, Phys.Rev.B **62**, 7477 (2000).
6. We say that S-2DEG boundary is "perfect" when the probability of Andreev reflection of an electron or hole quasiparticle with zero energy and momentum directed perpendicular to the surface is close to unity.
7. L.I.Glazman, G.B.Lesovik, D.E.Khmelnitskii *et al.,* Pisma Zh.Eksp.Teor.Fiz. **48**, 218 (1998)[JETP Lett. **48**, 238 (1988)].
8. C.J.Lambert, J.Phys.:Condens.Matter **3**, 6579(1991); Y.Takane and H.Ebisawa, J.Phys.Soc.Jpn. **61**, 1685 (1992).
9. M.Born, E.Wolf, "Principles of optics", Pergamon Press, 1986, p. 341.
10. H.Baranger, D.DiVincentzo, R.Jalabert, *et al*, Phys.Rev.B **44**, 10637 (1991).
    K.Richter, "Semiclassical theory of mesoscopic quantum systems", Springer-Verlag Berlin Heidelberg, 2000, (Springer tracts in modern physics; Vol. 161), pp. 63-68.
11. G.E.Blonder, M.Tinkham, and T.M.Klapwijk, Phys.Rev.B **25**, 4515 (1982).
12. I.S.Gradshtein, I.M.Ryzhik, "Table of integrals, series, and products", Academic Press, Inc, 1980.